\documentstyle[12pt]{article}
\newcommand{\ket}[1]{\left | #1 \right \rangle}

\newcommand{\Z}[1]{{\cal Z}_{#1}}
\begin{document}
\noindent
{\small Submitted to {\em Proc. Roy. Soc. Lond. A} for the Proceedings
of the Santa Barbara Conference on {\em Quantum Coherence and Decoherence}
held in December 1996, edited by E. Knill, R. LaFlamme and W. Zurek.}

\begin{center}
{\large\bf Quantum Algorithms and the Fourier Transform} \bigskip \\
Richard Jozsa\\ School of Mathematics and
Statistics\\ University of Plymouth\\ Plymouth, Devon PL4 8AA, England.
\\ Email: rjozsa@plymouth.ac.uk\\
\end{center}
\bigskip
{\bf Abstract:}
{\em The quantum algorithms of Deutsch, Simon and Shor are described in a way
which highlights their dependence on the Fourier transform.
The general construction of the Fourier transform on an Abelian
group is outlined and this provides a unified way of understanding 
the efficacy of the algorithms. Finally we describe an efficient quantum
factoring algorithm based on a general formalism of Kitaev and contrast
its structure to the ingredients of Shor's algorithm.
}\\[5mm]
{\large\bf   Introduction}

The principal quantum algorithms which provide an exponential speedup
over any known classical algorithms for the corresponding problems
are Deutsch's algorithm \cite{DJ}, Simon's algorithm \cite{SI} and
Shor's algorithm \cite{SH}. Each of these rests essentially
on the application of a suitable Fourier transform.
In this paper we will outline the construction of the Fourier transform
over a general (finite) Abelian group and highlight its origin and
utility in the quantum algorithms. This provides a unified way
of understanding the special efficacy of these algorithms.
Indeed we have described elsewhere \cite{J1} how this efficacy 
may be explicitly seen as a property of quantum entanglement 
in the context of implementing the large unitary operation which
is the Fourier transform.

From our general group-theoretic viewpoint we will see that Simon's 
and Shor's algorithms are essentially identical in their basic formal
structure differing only in the choice of underlying group. Both
algorithms amount to the extraction of a periodicity relative to
an Abelian group $G$ using the Fourier transform of $G$ in a uniform
way. This general viewpoint may also be useful in developing new
quantum algorithms by applying the formalism to other groups.

Kitaev \cite{KIT} has recently formulated a group--theoretic approach
to quantum algorithms. We will describe below a special explicit case
of his general formalism -- an efficient quantum factoring algorithm
which appears to be quite different from Shor's.
In particular, the Fourier transform as such, is not explicitly used.
It is especially interesting to contrast (rather than align!) Shor's
and Kitaev's algorithms as this may provide a new method -- in
addition to the ubiquitous Fourier transform -- for constructing quantum 
algorithms. 
The quantum searching algorithm of Grover \cite{GR} 
is also based on the Fourier transform but is of a different character 
from those mentioned above and we will not discuss it here.\\[5mm]
{\large\bf Some Notation}

We will write $B =\{0,1\}$ for the additive group of integers mod 2
and denote by $\cal B$ the Hilbert space of one qubit (i.e. a 2 dimensional
Hilbert space) equipped with a standard basis denoted by
$\{ \ket{0}, \ket{1} \} $. ${\cal B}^n$ will denote the Hilbert space
of $n$ qubits. The dual basis of $\cal B$ denoted by
$ \{ \ket{0'}, \ket{1'} \} $
is defined by
\begin{equation}
\ket{0'} = \frac{1}{\sqrt{2}} \left( \ket{0}+\ket{1}\right)
\hspace{1cm} \ket{1'} = \frac{1}{\sqrt{2}} \left( \ket{0}-\ket{1} 
\right)
\end{equation}
$H$ will denote the fundamental unitary matrix
\begin{equation}\label{h}
 H=  \frac{1}{\sqrt{2}}\left( \begin{array}{cc}
1 & 1 \\
1 & -1
\end{array} \right)   \end{equation}
Thus $H^2 =I$ and $H$ interchanges the standard and dual bases.
In terms of real geometry the dual basis lies on the $45^\circ$ lines
between the orthogonal directions $\ket{0}$ and $\ket{1}$ and $H$ is
the transformation given by reflection in a line at angle $\pi/8$
to the $\ket{0}$ direction. Thus the eigenvectors of $H$ 
(parallel and perpendicular to the mirror line) are
$\cos \frac{\pi}{8} \ket{0} \pm \sin \frac{\pi}{8} \ket{1}$ 
belonging to $\lambda = \pm 1$
respectively. We will see later that $H$ is also the Fourier transform
on the group $B$. 

The elements of $B^n$ are $n$ bit strings.
If $x=(x_1 , \ldots ,x_n)$ and $y=(y_1 , \ldots ,y_n)$ are in $B^n$
then we write
\[ x\oplus y = (x_1 \oplus y_1 , \ldots , x_n \oplus y_n ) \in B^n \]
\[ x\cdot y = (x_1 y_1 \oplus \cdots \oplus x_n y_n ) \in B \]
(the operations on the RHS's being addition and multiplication mod 2 in $B$.)
Note that $x\cdot y$ is the parity of the number of places where $x$ and $y$
both have a bit value of 1.\\[5mm]
{\large\bf Early Days}

The earliest quantum algorithms \cite{DD1,DJ} were concerned with a
situation in which we are given a ``black box'' or oracle that computes
a function $f:B^n \rightarrow B$
and we are required to decide whether a certain ``global'' property
(i.e. a joint property of all the function values) holds of $f$.
For quantum computation the black box is given as
a unitary transformation ${\cal U}_f$ on $n+1$ qubits given in the
standard basis by
\begin{equation} \label{uf}
 {\cal U}_f:\,\,
\underbrace{\ket{x_1}\ket{x_2} \ldots \ket{x_n}}_{\rm input}
\ket{y} \longrightarrow
   \ket{x_1}\ket{x_2} \ldots \ket{x_n}\ket{
y \oplus f(x_1 ,\ldots ,x_n )} \end{equation}
(We will often abbreviate $\ket{x_1}\ket{x_2} \ldots \ket{x_n}$ as
$\ket{x}$ for $x\in B^n$.) Thus if $y$ is initially set to 0
the value of $f$ may be read from the last qubit.

For our first problem, referred to as Deutsch's XOR problem \cite{DD1}, 
we have $n=1$ so that $f$ is one of the four possible functions
$f:B\rightarrow B$. We are to decide whether $f(0)\oplus f(1)$ is
0 or 1. Equivalently we wish to decide whether $f$ is a constant
function or a ``balanced'' function (where balanced means that 
$f$ takes one value 0 and one value 1). Clearly any classical
computer requires evaluating $f$ twice to decide this. According to
Deutsch's original method \cite{DD1}, 
the problem may be solved on a quantum computer
after running ${\cal U}_f$ only {\em once}
but the algorithm succeeds only with probability $\frac{1}{2}$ (and we
know when it has been successful). The method is simply to
run ${\cal U}_f$ on the input superposition $\frac{1}{\sqrt{2}} (
\ket{0}+\ket{1})$ yielding the state $\frac{1}{\sqrt{2}}(
\ket{0}\ket{f(0)} + \ket{1} \ket{f(1)})$. Writing this state in the
dual basis we have the four possibilities
given by the two constant functions:
\[ \frac{1}{\sqrt{2}} \left( \ket{0}\ket{f(0)}+\ket{1}\ket{f(1)} \right) =
\left\{ \begin{array}{ccc} 
\frac{1}{\sqrt{2}}(\ket{0}\ket{0}+\ket{1}\ket{0}) & = & 
\frac{1}{\sqrt{2}}(
\ket{0'}\ket{0'} + \ket{0'}\ket{1'})\\
\frac{1}{\sqrt{2}}(\ket{0}\ket{1}+\ket{1}\ket{1}) & = & 
\frac{1}{\sqrt{2}}(
\ket{0'}\ket{0'} - \ket{0'}\ket{1'})
\end{array} \right.   \]
and the two balanced functions:
\[ \frac{1}{\sqrt{2}} \left( \ket{0}\ket{f(0)}+\ket{1}\ket{f(1)} \right) = 
\left\{ \begin{array}{ccc} 
\frac{1}{\sqrt{2}}(\ket{0}\ket{0}+\ket{1}\ket{1}) & = &
\frac{1}{\sqrt{2}}(
\ket{0'}\ket{0'} + \ket{1'}\ket{1'})\\
\frac{1}{\sqrt{2}}(\ket{0}\ket{1}+\ket{1}\ket{0})  & = &
\frac{1}{\sqrt{2}}(
\ket{0'}\ket{0'} - \ket{1'}\ket{1'})
\end{array} \right.
\]  
Now measure the second qubit in the dual basis. If the result is
$0'$ (which occurs with probability $\frac{1}{2}$  
in every case) then we have lost all the information about the function $f$.
If the result is $1'$ then measurement of the first qubit
will reliably distinguish between constant and balanced functions.

In our second algorithm \cite{DJ}, 
referred to as Deutsch's algorithm, we are given 
$n$ and a function $f:B^n \rightarrow B$. It is promised that $f$ is either 
constant or balanced (where balanced means that $f$ takes values 0 and 1 an
equal number of times i.e. $2^{n-1}$ times each). The problem is to decide 
whether $f$ is balanced or constant. The method, described in detail in 
\cite{DJ}, involves running ${\cal U}_f$ {\em twice} (and using $H$ $O(n)$
times) to construct the state
\begin{equation} \label{f}
\ket{f} = \frac{1}{\sqrt{2^n}} \sum_{x\in B^n} (-1)^{f(x)} \ket{x} 
\end{equation}
Then $\ket{f}$ for any constant function is orthogonal to 
the corresponding state
for any balanced function and thus 
we can solve our decision problem with certainty
by a suitable measurement on the resulting state. The quantum algorithm
always runs in time $O(n)$ whereas any classical algorithm (which gives
the result with certainty in every case) will require time of $O(2^n )$
at least in some cases.

Note that Deutsch's XOR problem is the $n=1$ case of the above decision 
problem. However the above algorithm, running ${\cal U}_f$ twice, offers 
no advantage over the obvious classical algorithm for $n=1$.
Another distinction between the above two algorithms is that
the XOR problem is solved only with probability 1/2 whereas
the second algorithm is {\em always} succesful. An interesting recent
innovation \cite{WHO} fully unifies and considerably improves the above 
two algorithms: the XOR problem may be solved with {\em certainty} and the
state in eq. (\ref{f}) may be constructed by running ${\cal U}_f$ 
only {\em once}. The improved XOR algorithm is then precisely the
$n=1$ case of the improved Deutsch algorithm. The basic idea is
to set the output register to the state $\frac{1}{\sqrt{2}}(
\ket{0}-\ket{1})$ before applying ${\cal U}_f$. Note that by eq. (\ref{uf})
\[ {\cal U}_f : \ket{x} (\ket{0}-\ket{1}) \longrightarrow
\left\{ \begin{array}{rl}
   \ket{x} (\ket{0}-\ket{1}) & \mbox{if $f(x)=0$} \\
  -\ket{x} (\ket{0}-\ket{1}) & \mbox{if $f(x)=1$}
\end{array} \right. \]
Thus
\[ {\cal U}_f : \frac{1}{\sqrt{2^n}} \sum_{x\in B^n } \ket{x}
\left( \frac{\ket{0}-\ket{1}}{\sqrt{2}} \right) \longrightarrow
\left( \frac{1}{\sqrt{2^n}} \sum_{x\in B^n } (-1)^{f(x)} \ket{x} \right)
\left( \frac{\ket{0}-\ket{1}}{\sqrt{2}} \right)  \]
giving the state $\ket{f}$ in the first $n$ qubits after only
one application of ${\cal U}_f$. The last qubit plays a curiously
passive role in that its state is unchanged in the process.
(This is reminiscent of the similarly passive role of the second
register in Shor's algorithm \cite{EJ,SH}).

The explicit description of the measurement on $\ket{f}$ which
distinguishes balanced from constant functions is significant for
subsequent developments. We first apply the operation $H$ to each 
of the $n$ qubits of $\ket{f}$. Denoting the resulting $n$-qubit
operation by $H_n$ we have, for each $x\in B^n$
\begin{equation}  \label{hn}
H_n : \ket{x} \rightarrow \frac{1}{\sqrt{2^n}}
\sum_{y\in B^n} (-1)^{x\cdot y} \ket{y} \end{equation}
Note that
\[ H_n \ket{0\ldots 0} = \frac{1}{\sqrt{2^n}} \sum_{y\in B^n} 
\ket{y} \]
is the equal superposition of all the standard basis states and
that up to an overall sign this coincides with $\ket{f}$ for $f$
constant. Since $H_n H_n =I$ it follows that $H_n \ket{f} = \ket{0\ldots 0}
$ for $f$ constant. Thus if $f$ is balanced then $H_n \ket{f}$ 
must be orthogonal to $ \ket{0\ldots 0}$ i.e. $\ket{f}$ lies in 
the span of $\{ \ket{x} : x\neq 0\ldots 0\} $.  Hence to distinguish
balanced from constant functions we apply $H_n$ to $\ket{f}$
and then read the bits to see whether they are all zero or not.

The above measurement has $2^n$ natural outcomes (i.e. all $n$-bit strings)
and we may ask if there are special balanced functions which yield
with certainty the other outcomes $x \in B^n$ in the same way that constant 
functions lead to the outcome $0\ldots 0$. For each $k\in B^n$ 
consider the function $f_k :B^n \rightarrow B$ given by
\[ f_k (x) = k\cdot x \]
It is easily verified that each $f_k$ is a balanced function
for $k\neq 0\ldots 0$ (giving a small subset of all possible balanced
functions). We will see later that the operation $H_n$ is the Fourier
transform on the additive group $B^n$ (also known as the Walsh or Hadamard 
transform) and the functions $f_k$ are the Fourier (Walsh, Hadamard)
basis functions.
For these functions we have
\[ H_n \ket{f_k} = \ket{k} \]
which follows readily by comparing eq. (\ref{f}) with eq. (\ref{hn})
and the fact that $H_n H_n =I$.
Thus our quantum algorithm can reliably distinguish the $2^n$
functions $f_k$ after evaluating the function only {\em once}! 
However this finer use of the measurement outcomes does not represent an 
exponential advantage over classical computation  
since the classical evaluation of just $n$ values of $f_k$ on the inputs
$10\ldots 0$, $010\ldots 0$, up to $0\ldots 01$ will successively reveal the 
$n$ bits of $k$. 

A significant feature of the problem of distinguishing balanced from constant
functions is the following: if we tolerate any (arbitrarily small) non-zero
probability of error in the result then we lose the exponential
advantage of the quantum algorithm over classical algorithms. Indeed given 
any $\epsilon$, if we sample $O(-\log \epsilon )$ random values of $f$
then we can determine within error probability $\epsilon$  
whether $f$ is balanced or constant by just claiming ``constant''
if all the sampled values are the same. However the 1 versus $n$ gap between 
the quantum and classical identification of $f_k$ described above  
persists even if we tolerate a small probability of error in the result.
This led Bernstein and Vazirani \cite{BV} to amplify this gap
to a super--polynomial size by a recursive procedure, leading to
the first example of a problem which could be solved exponentially faster
by a quantum algorithm than by any classical algorithm even if a small
 probability of error is tolerated. Soon thereafter Simon \cite{SI}
gave a simpler example. Below we will describe the structure of Simon's 
algorithm and Shor's algorithm emphasising their similarity, which will
lead naturally to the general concept of the Fourier transform on an 
Abelian group. \\[5mm]
{\large\bf Simon's Algorithm}

We are given a ``black box'' (or oracle) which computes a function
$f:B^n \rightarrow B^n$. The function is promised to be a 2-to-1
function and have periodicity $\xi \in B^n$ i.e.
\begin{equation}  \label{per1}
f(x)=f(y) \hspace{5mm}\mbox{ iff } y=x\oplus \xi \hspace{5mm}
\mbox{ for all } x,y \in B^n
\end{equation}
Our problem is to find $\xi$ efficiently (i.e. in poly($n$) steps,
each evaluation of the function counting as one step).
More precisely, the function is given as a unitary transformation
$U_f$ on ${\cal B}^{2n}$ defined by
\[ U_f : \ket{x}\ket{y} \rightarrow \ket{x}\ket{y\oplus f(x)}. \]
Simon's algorithm (omitting normalisation factors) is the following:
\begin{description}
\item[Step 1.] Start with the state $\ket{0\ldots 0}\in {\cal B}^n$
and apply $H_n$ to get $\sum_{x} \ket{x}$.
\item[Step 2.] Apply $U_f$ to $(\sum\ket{x})\ket{0}$ to get
$\sum\ket{x}\ket{f(x)}$.
\item[Step 3.] Measure the value of register 2 and keep the corresponding
state of register 1. By eq. (\ref{per1}) the state of register 1 will
have the form $\ket{x_0} + \ket{x_0 \oplus \xi}$
where $x_0 \in B^n$ has been chosen equiprobably.
\item[Remark.] Thus we have set up a state involving a periodic
superposition of $\ket{x_0}$ and $\ket{x_0 \oplus \xi}$
(noting that $x_0 \oplus \xi \oplus \xi = x_0$ etc.)
This contains the desired information of $\xi$ together with
an unwanted randomly chosen $x_0$. A direct measurement of the label would 
yield any $x\in B^n$ equiprobably, providing no information at all about
$\xi$.
\item[Step 4.] Apply $H_n$ to get (c.f. eq. (\ref{hn}))
\[ \sum_{y\in B^n} \left( (-1)^{x_0\cdot y} + (-1)^{(x_0\oplus \xi )\cdot y}
\right)  \ket{y} = \pm\sum_{y:\, y\cdot \xi =0} \ket{y} \]
(where the overall sign depends on $x_0$). Note that if $y\cdot\xi = 1$
then the terms on the LHS will interfere destructively.
\item[Remark.]  The effect of $H_n$ here is to
wash out the unwanted $x_0$ from the labels
  and to invert the information of $\xi$,
recoding it as $y$ such that $y\cdot \xi =0$. 
A direct measurement of the label will now yield information about $\xi$.
The same formal features
will arise in Shor's algorithm below.
\item[Step 5.] Measure the register to find a value of $y$
(equiprobably) such that $y\cdot \xi = 0$.
\item[Step 6.] Repeat the above to find enough $y_i$'s so that
$\xi$ may be determined by solving the linear system
$y_1 \cdot \xi = 0,\ldots  ,y_k \cdot \xi = 0$ .
It may be shown that $O(n^2 )$ repetitions suffice to determine
$\xi$ with any prescribed probability $p<1$.
\end{description}
{\large\bf Shor's Algorithm}

Shor's algorithm for factoring a given number $N$ \cite{SH,EJ}
proceeds by solving an equivalent problem: given any $y$ coprime to $N$
find the order $r$ of $y$ mod $N$. (Note that if $y\leq N$ is
chosen {\em at random}
then we may use Euclid's algorithm \cite{EJ} to efficiently
determine whether $y$ is coprime to $N$ or not. If it is not coprime, then the
highest common factor of $y$ and $N$ gives a factor of $N$ directly.)
The order $r$ of $y$ mod $N$ is the least integer $r$ such that
\[ y^r \equiv 1 \mbox{ mod } N \]
Let $\Z{n}$ denote the group of integers mod $n$. For any $q$ we have a
function
\[ f: \Z{q} \rightarrow \Z{N} \]
\[ f(x) = y^x \mbox{ mod } N \]
so that
\begin{equation}\label{per2}
f(x+r) = f(x) \hspace{1cm} \mbox{ if } \hspace{1cm} x+r \leq q
\end{equation}
Note that because of the condition $x+r \leq q$,  this function is not
wholly periodic on $\Z{q}$ unless $q$ is an exact multiple of
(the unknown) $r$. However if $q$ is chosen sufficiently large, then
the slight spoiling of the periodicity at $x$ near $q$ 
(i.e. in one period only) will have
a negligible effect. Ideally we would choose $q=\infty$ here
for perfect periodicity in every case but in practice we require that
$q$ be finite.

Thus Shor's algorithm combines two separate issues: firstly the extraction
of the periodicity of $f$ and secondly, dealing with the fact that $f$ is 
not {\em perfectly} periodic. In our description below we will focus
on the first issue and assume for simplicity
that $q$ is an exact multiple of $r$.
We will discuss this assumption and the second issue at the end.

Suppose we are given a fixed $y$ coprime to $N$ and we
want to compute its order mod $N$. The unitary transformation
\[ U_f : \ket{x_1}\ket{x_2} \rightarrow \ket{x_1}\ket{x_2 + y^{x_1}
\mbox{ mod } N}  \hspace{1cm} x_1 \in \Z{q} \hspace{8mm}
x_2 \in \Z{N} \]
is efficiently computable \cite{SH,EJ} and will play the same role as
$U_f$ in Simon's algorithm.
Shor's algorithm proceeds by the following steps which
parallel exactly the steps of Simon's algorithm. $DFT_q$ below denotes
the discrete Fourier transform for integers mod $q$.
It is defined by
\begin{equation} \label{dft}
DFT_q : \ket{k} \rightarrow \frac{1}{\sqrt{q}}
\sum_{l=0}^{q-1} e^{2\pi i\frac{ kl}{q}}
\ket{l} \hspace{1cm} k\in \Z{q}
\end{equation}
and replaces $H_n$ in Simon's algorithm. As before we will omit normalisation
factors.
\begin{description}
\item[Step 1.] Start with the state $\ket{0}$ (in a $q$ dimensional
Hilbert space) and apply $DFT_q$ to get $\sum_{x=0}^{q} \ket{x}$.
\item[Step 2.] Apply $U_f$ to $(\sum \ket{x})\ket{0}$ to get
$\sum \ket{x}\ket{y^x \mbox{ mod } N}$.
\item[Step 3.] Measure the value of register 2 and keep the corresponding
state of
register 1. This state will have the form
 $\sum_{\lambda} \ket{x_0 + \lambda r \mbox{ mod } q}$,
where $x_0 \in \Z{r}$ has been
chosen equiprobably.
\item[Remark.] As in Simon's algorithm a direct measurement of the label
will give no information at all about $r$. 
\item[Step 4.] Apply $DFT_q$. Using eq. (\ref{dft}) we get \cite{EJ}
 a state of the form
\[ \sum_{k\in \Z{r}} e^{i\phi_k (x_0 )}\ket{k\frac{q}{r}} \]
\item[Remark.] Note that as in Simon's algorithm the random shift $x_0$ 
has been eliminated from the labels and the information of $r$ has been
inverted as $kq/r$.
\item[Step 5.] Measure the register to
get a multiple
$c=k(q/r)$ where $k \in \Z{r}$ has been chosen equiprobably.
Thus $c/q=k/r$ where $c$ and $q$ are known. 
\item[Step 6.] Repeat the above until we get a result corresponding to
$k$ being coprime to $r$. Then $r$ is obtained by cancelling $c/q$
down to its lowest terms. It may be shown \cite{SH,EJ}
that $O(\log N)$ repetitions will suffice to determine $r$ with any prescribed
probability $p<1$.
\end{description}
Thus we see that Simon's and Shor's algorithms are structurally
identical (in the ideal case that $q$ is an exact multiple of $r$
or $q=\infty$). The group $B^n$ and the operation $H_n$ have been replaced
respectively by the group $\Z{q}$ and operation $DFT_q$. 
We will see in the next section that
these operations are just the Fourier transforms for the respective Abelian
groups and the general construction of the Fourier transform will
clarify their role in the preceeding algorithms.

In general $q$ cannot be guaranteed to be a multiple of $r$. Let 
us write $q=Kr+a$ with $a<r<N$ and let $q_0 = Kr$.
In step 3 of the algorithm, instead of
\[ \ket{\psi_{q_0}} = \frac{1}{\sqrt{K}} \sum_{\lambda =0}^{K-1}
\ket{x_0 + \lambda r} \]
we will get
 \[ \ket{\psi_{q}} = \frac{1}{\sqrt{K+1}} \sum_{\lambda =0}^{K}
\ket{x_0 + \lambda r} \]
possibly containing at most one extra term (as written)
if $x_0 < a$.
Thus for sufficiently large $K$, $\ket{\psi_{q_0}}$ and $\ket{\psi_{q}}$
may be as close as desired. In step 4 we will apply
$DFT_q$ to $\ket{\psi_{q}}$ rather than $DFT_{q_0}$ to $\ket{\psi_{q_0}}$.
However $q-q_0 =a<N$ so if $q$ is chosen sufficiently large
compared to $N$ we may expect that the two actions will result in
close outcomes. In step 5 $c$ will not be an exact multiple of $q/r$
but will be near to such a multiple with high probability.
These intuitive remarks may be formalised \cite{SH,EJ} to  
show that a choice of $q$ of order $N^2$ suffices
determine $r$. In step 5 the fraction $k/r$ 
is then uniquely determined from the suitably close 
rational approximation $c/q$ by using the theory of
continued fractions \cite{EJ}.\\[3mm]
{\large\bf The Fourier Transform on an Abelian Group}

Let $G$ be a (finite) Abelian group and let $\cal H$ be a Hilbert space
with an orthonormal basis $\{ \ket{g}: g\in G \}$ (the ``standard'' basis)
labelled by the elements of $G$. There is a natural unitary shifting action
of $G$ on $\cal H$ given by
\begin{equation} \label{sh}
h: \ket{g} \rightarrow \ket{hg} \hspace{1cm} h,g \in G
\end{equation}
Note that we use multiplicative notation for the operation in the group
$G$ and we use the same symbol (e.g. $h$ in eq. (\ref{sh}) above)
to denote a group element and its unitary action on $\cal H$.

Let $f:G\rightarrow X$ be a function on the group (taking values in some
set $X$) and consider
\[ K= \{ k\in G : f(kg) = f(g) \mbox{ for all $g\in G$} \} \]
$K$ is necessarily a subgroup of $G$ called the stabiliser or
symmetry group of $f$. It characterises the periodicity of $f$
with respect to the group operation of $G$. Given a device that computes 
$f$, our aim is to determine $K$. More precisely we wish to determine
$K$ in time $O({\rm poly}(\log |G|))$ where $|G|$ is the size of $G$ and
the evaluation of $f$ on an input counts as one computational step.
(Note that we may easily determine $K$ in time $O({\rm poly}( |G|))$ by simply
evaluating and examining all the values of $f$). 
Further discussion of this time constraint will be given in the next section.

We begin by constructing the state
\[ \ket{f}= \frac{1}{\sqrt{|G|}} \sum_{g\in G} \ket{g}\ket{f(g)}  \]
and read the second register. Assuming that $f$ is suitably non-degenerate
-- in the sense that $f(g_1 ) = f(g_2 )$ iff $g_1 g_{2}^{-1} \in K$ i.e.
that $f$ is one-to-one within each period -- we will obtain in the first
register
\begin{equation} \label{per}
\ket{\psi (g_0 )} = \frac{1}{\sqrt{|K|}} \sum_{k\in K} \ket{g_0 k}
\end{equation}
corresponding to seeing $f(g_0 )$ in the second register
and $g_0$ has been chosen at random.\\[3mm]
{\em Examples.} In Simon's algorithm $G$ is the additive group $B^n$
and $K$ is the cyclic subgroup $\{0,\xi \}$ generated by $\xi$.
In Shor's algorithm $G$ is the additive group $\Z{q}$ and $K$ is
the cyclic subgroup $\{ 0,r,2r, \ldots \}$ generated by $r$.
In each case $K$ is specified by giving its generator. The state
(\ref{per}) is obtained in step 3 of the algorithm. $\Box$\\[3mm]
{\em Remark.} The construction leading to the state (\ref{per})
applies in a more general context than just a function on a group.
Suppose we have any mathematical object $F$ with an action of the
group $G$ on it:
\[ g:F \rightarrow gF \hspace{1cm} \mbox{ such that }
(g_1 g_2 )F = g_1 (g_2 F). \]
The symmetry group of $F$ is the subgroup $K = \{ k\in G: kF=F \}$.
By constructing $\sum_g \ket{g}$, applying it to a suitable state 
description $\ket{F}$ of $F$ and reading the second register we 
obtain the state $\sum_k \ket{g_0 k}$ as in eq. (\ref{per}). $\Box$

In eq. (\ref{per}) we have an equal superposition of labels corresponding
to a randomly chosen coset of $K$ in $G$. Now $G$ is the disjoint
union of all the cosets so that if we read the label in eq. (\ref{per})
we will see a random element chosen equiprobably from all of $G$
yielding no information at all about $K$.
The Fourier transform will provide a way of eliminating $g_0$
from the labels which may then provide direct information about $K$.
We first construct a basis $\ket{\chi_i}$ of states which are
{\em shift invariant} in the sense:
   \[ g\ket{\chi_i} = e^{\phi_{i}(g)} \ket{\chi_i} \mbox{ for all $g\in G$} \]
Such states are guaranteed to exist since the shift operations $g$
are unitary and they all commute. Next note that the state in 
eq. (\ref{per}) may be written as a $g_0$-shifted state:
\[ \sum_{k\in K} \ket{g_0 k} = g_0 \left( \sum_{k\in K} \ket{k} \right) \]
Hence if we write this state in the basis $\{ \ket{\chi_i}, i=1,\ldots ,|G|\}$
then $\sum_k \ket{k}$ and $\sum_k \ket{g_0 k}$ will contain the same 
pattern of labels, determined by the subgroup $K$ only.
The Fourier transform is simply defined to be the unitary operation
which transforms  the shift-invariant basis into the standard basis.
After applying it to eq. (\ref{per}) we may read the shift-invariant basis 
label by reading in the standard basis. This explains the essential role
of the Fourier transform in step 4 of the algorithms.

The shift-invariant states $\ket{\chi_i}$ are constructed using some basic 
group representation theory \cite{FUL}. Consider any (nonzero) complex valued 
function on the group
\[ \chi : G\rightarrow {\cal C} \]
which respects the group operation in the sense that
\begin{equation} \label{mult}
\chi (g_1 g_2 ) = \chi (g_1 ) \chi (g_2 )  
\end{equation}
For Abelian groups these are the irreducible representations
\cite{FUL} of $G$.
By listing the values $\chi$ may also be viewed as a complex vector of
dimension $|G|$.

For our purposes the essential properties of these functions are
the following (c.f. \cite{FUL} for a full discussion and proofs).
\begin{description}
\item[(A)] Any value $\chi (g)$ is a $|G|^{\rm th}$ root of unity.
\item[(B)] Orthogonality (Schur's lemma): For each $i$ and $j$
\begin{equation} \label{ortho}
\frac{1}{|G|} \sum_{g \in G} \chi_i (g) \overline{\chi_j (g)} = \delta_{ij} 
\end{equation}
(where the overline denotes complex conjugation).
\item[(C)] There are always exactly $|G|$ different functions $\chi$ 
satisfying eq. (\ref{mult}).
\end{description}
It is remarkable that the simple condition eq. (\ref{mult}) has such
strong consequences. In particular the orthogonality condition
{\bf (B)} entails the fact that the Fourier transform as a linear
transformation is {\em unitary} rather than just invertible.
This appears to make no significant difference for classical
computation but it is crucial for quantum computation!    

Since {\bf (B)} provides the fundamental connection to quantum computation
we give a simple proof of it (incorporating also {\bf (A)}).
Note that by (\ref{mult})
$\chi (e)=1$ where $e$ is the identity of $G$. Also (by Lagrange's theorem)
we have
$g^{|G|} = e$ for all $g\in G$. Hence $\chi (g)$ is always a $|G|^{\rm th}$
root of unity so $\overline{\chi (g)} = \chi (g^{-1})$. Now for any
$\chi_1$, $\chi_2$ consider:
\[ \chi_1 (h) \left( \sum_{g\in G} \chi_1 (g) \chi_2 (g^{-1}) \right)
= \sum_{g\in G} \chi_1 (hg) \chi_2 (g^{-1}) \]
\begin{equation} \label{hg} = \sum_{\tilde{g}\in G} \chi_1 (\tilde{g}) 
\chi_2 (\tilde{g}^{-1}h) \hspace{1cm}
\mbox{(putting } \tilde{g}=hg ) \end{equation}
\[ = \left( \sum_{\tilde{g}} \chi_1 (\tilde{g}) \chi_2
(\tilde{g}^{-1})\right) \chi_2 (h) \]
Hence for every $h\in G$
\[ \left( \chi_1 (h) - \chi_2 (h)\right) \sum_{g\in G} \chi_1 (g)
\overline{\chi_2 (g)} =0 \]
giving orthogonality if $\chi_1 \neq \chi_2 $.
If $\chi_1 = \chi_2 = \chi $ then
\[ \sum_g \chi (g) \overline{\chi (g)} = \sum \chi (g) \chi (g^{-1})
= \sum \chi (e) = \sum 1 = |G| \]
completing the proof of (\ref{ortho}).

For any function $\chi_i$ satisfying eq. (\ref{mult}) consider the state
\[ \ket{\chi_i} = \frac{1}{\sqrt{|G|}}\sum_{g\in G} 
\overline{\chi_i (g)} \ket{g} \]
The orthogonality relation (\ref{ortho}) implies that the states
$\{ \ket{\chi_i}: i=1, \ldots ,|G|\}$ form an orthonormal
basis of $\cal H$, called the Fourier basis. 
Furthermore these basis states are shift-invariant in the required sense:
\begin{equation}\label{shift}
h\ket{\chi_i} = \chi_i (h) \ket{\chi_i}\hspace{1cm} h\in G
\end{equation}
which is easily verified using eqs. (\ref{sh}), (\ref{mult})
and making the same replacement as in eq. (\ref{hg}).

Let us choose an ordering $g_1 ,g_2 , \ldots ,g_{|G|}$ of the elements
of $G$. The Fourier transform $FT$ on $G$ (with respect to the ordering)
is defined to be the unitary transformation which maps $\ket{\chi_i}$
to $\ket{g_i}$. Thus in the ordered basis $\{ \ket{g_i} \}$ the matrix of
$FT$ is formed by listing the values of the functions $\chi_i$ as rows:
\begin{equation} \label{ft}
[FT]_{ij} = \frac{1}{\sqrt{|G|}} \chi_i (g_j ) 
\end{equation}
{\em Examples}.
If $G=\Z{q}$ then the $q$ functions $\chi_k$ are defined by
\[ \chi_k (1) = e^{2\pi i k/q} \hspace{1cm} k=0, \ldots , q-1 \]
and by (\ref{mult}) $\chi_k (m) = \chi_k (1)^m
= \exp 2\pi i\, km/q$ for all $m\in \Z{q}$.
These values scaled by $\sqrt{q}$ are the rows of the matrix of $DFT_q$.\\
For $G = B^n$ the $2^n$ $\chi$ functions are
\[ \chi_{\sigma}(x) = (-1)^{x\cdot \sigma} \hspace{1cm}
\mbox{for all } x, \sigma \in B^n \]
which (scaled by $\sqrt{2^n}$) are the rows of the Hadamard transform $H_n$
(c.f. eq. (\ref{hn})).\\[3mm]
{\large\bf Efficient Computation of the Fourier Transform}

The Fourier transform $FT$ on $G$ is a unitary operation of size $|G|$.
It is known \cite{DD1,EJ} that any unitary operation of size $d$ 
may be implemented in time $O(d^2 )$ but this does not suffice for our
application of $FT$. In Simon's algorithm $|G|=2^n$ but we want the algorithm 
to run in poly($n$) time and in Shor's algorithm $|G|=O(q)=O(N^2 )$
and we want the algorithm to run in poly($\log N$) time. Thus we want to
implement $FT$ in poly($\log |G| $) time.

In classical computation the application of a matrix   
of size $|G|$ requires time $O(|G|^2 )$. The classical fast Fourier transform
($FFT$) algorithm (applicable to certain groups) improves this to
$O(|G|\log |G| )$ but this, in itself, does not suffice for our
quantum algorithms since it is still exponetial in $\log |G|$.
It may be seen that in a quantum context the implementation of
the $FFT$ algorithm combines with extra non-classical properties of
entanglement to provide an algorithm which runs in $O({\rm poly}(\log |G|))$
time. This feature has been elaborated in \cite{J1} and is also
discussed in \cite{HO}.\\[3mm]
{\large\bf Kitaev's Algorithm}

An approach to the construction of quantum algorithms based on
group-theoretic principles (for Abelian groups) has recently been
developed by Kitaev \cite{KIT}. We describe here
an explicit example of his general formalism -- 
an alternative efficient quantum factoring algorithm.
This algorithm, in contrast to Shor's, does not explicitly require the Fourier
transform to be performed and appears to be based on different principles.

Kitaev's algorithm, like Shor's, proceeds by finding the order $r$
of a number $y$ coprime to $N$. Let $U: {\cal H}_N \rightarrow
{\cal H}_N$ be the unitary operator on an $N$ dimensional Hilbert space
given by ``multiplication by $y$'' (easily implementable efficiently):
\begin{equation} \label{um}
 U: \ket{m} \mapsto \ket{my \mbox{ mod } N} \hspace{1cm}
m=0, \ldots ,N-1 \end{equation}
Thus we will be focussing on the {\em multiplicative} structure
of the integers mod $N$ (rather than the additive structure)
and working in a Hilbert space of dimension $N$. We do not need to
choose a $q\approx O(N^2 )$ as in Shor's algorithm and the associated
complications of $q$ not being an exact multiple of $r$ do not arise.

Since $U^r =I$ we see that the eigenvalues of $U$ are $r^{\rm th}$ roots of
unity i.e. $\lambda_k = \exp{(-2\pi ik/r)}, k=0, \ldots ,r-1$.
It is straightforward to verify that the following states
$\ket{\lambda_k}$ are eigenstates of $U$ belonging respectively 
to the eigenvalues $\lambda_k$:
\begin{equation}\label{eig}
\ket{\lambda_k} = \frac{1}{\sqrt{r}}\sum_{l=0}^{r-1}
\exp{(2\pi i\frac{lk}{r})} \ket{y^l \mbox{ mod } N} \hspace{1cm} k=0, \ldots , r-1
\end{equation}
 and that
\begin{equation}\label{sum}
\ket{1}= \frac{1}{\sqrt{r}}\sum_{k=0}^{r-1} \ket{\lambda_k}
\end{equation}
{\em Remark}. The fact that (\ref{eig}) are eigenstates of $U$ is
closely related to our previous construction of shift invariant
states. Indeed the multiplicative group of powers of $y$ mod $N$
is isomorphic to the additive group $\Z{r}$ (where we associate
$y^l$ with $l\in \Z{r}$). Under this isomorphism the operation $U$
becomes the shift operation of ``adding 1'' in $\Z{r}$.
Then (\ref{eig}) gives precisely the shift invariant states of $\Z{r}$
but written with multiplicative labels $y^l \mbox{ mod } N$ rather than
the additive labels $l\in \Z{r}$.\\
Eq. (\ref{sum}) is simply derived by noting that each
$\ket{\lambda_k }$ in (\ref{eig}) contains $\ket{1}$ with
amplitude $1/\sqrt{r}$. Hence the sum in (\ref{sum}) contains $\ket{1}$
with amplitude 1 so that all other $\ket{k}$'s with $k\neq 1$ 
must have amplitude 0 as (\ref{sum}) is a normalised state.
This equation also has a group-theoretic origin. It may be shown \cite{FUL}
that for any group $G$ if we sum all the $\chi_i$ functions we get:
\[ \frac{1}{|G|} \sum_{i=1}^{|G|} \chi_i (g) =
\left\{ \begin{array}{l}
1 \mbox{ if $g=e$} \\
0 \mbox{ if $g\neq e$}
\end{array} \right.
\]
Then (\ref{sum}) follows immediately using the above interpretation
of $\ket{\lambda_k }$ as shift invariant states.$\Box$ 

{\em Suppose} now that we have an efficient procedure for measuring the
eigenvalues of a unitary operator. More precisely, given a quantum
device which computes an $n$-qubit operation
$U$ and an eigenstate $\ket{\lambda}$
of $U$, suppose that we can compute the value of $\lambda$ efficiently
i.e. in time $O({\rm poly}(\log n))$.
Suppose furthermore that on an input {\em superposition} of eigenstates
$\sum a_k \ket{\lambda_k }$ the procedure returns some one of the eigenvalues
$\lambda_k$ with probability $|a_k |^2$.
Then applying this procedure to $U$ and the state $\ket{1}$ above,
we will be able to efficiently find a value of $k/r$ chosen
equiprobably for $k=0,\ldots , r-1$. As in Shor's analysis, this suffices to
factor $N$ efficiently. It is remarkable that the apparently humdrum
state $\ket{1}$ (when viewed appropriately as in eq. (\ref{sum}))
contains the information to factorise {\em any} given number!
\\[3mm]
{\large\bf How to Measure the Eigenvalues of $U$}{\large \,\cite{KIT}}\\[3mm]
Suppose we are given a ``black box'' which computes $U:
{\cal B}^n \rightarrow {\cal B}^n$ , a unitary operation on $n$ qubits,
and also
an eigenstate $\ket{\lambda}$ of $U$ with $\lambda = \exp{2\pi i \phi}$.
We want to measure $\phi$. The basic idea is to set up a state
$\ket{\alpha}=\sqrt{p_0}\ket{0}+\sqrt{p_1}\ket{1}$
whose amplitudes depend on $\phi$.
Then by sufficiently many measurements on copies of
$\ket{\alpha}$ we can estimate the probabilities $p_0 , p_1$ and
hence $\phi$.

We first describe how to implement $\Lambda (U)$, the ``controlled-$U$''
operation on $n+1$ qubits (which includes one ``control qubit'').\\
Let $\tau : {\cal B}^{2n}\rightarrow {\cal B}^{2n}$ on two $n$-qubit
registers ${\cal X}, {\cal Y}$ be the addition of $n$-bit strings:
\[ \tau : \ket{x}\ket{y} \mapsto \ket{x}\ket{x\oplus y}
\hspace{1cm} x,y \in {\cal B}^n \]
Let $\Lambda (\tau ) :{\cal B}^{2n+1}\rightarrow {\cal B}^{2n+1}$,
on a 1-qubit control register $\cal C$ with ${\cal X}$ and $\cal Y$,
be the controlled $\tau$ operation:
\[ \Lambda (\tau ) : \ket{0}\ket{x}\ket{y} \mapsto \ket{0}\ket{x}\ket{y}
\hspace{1cm}
\Lambda (\tau ) : \ket{1}\ket{x}\ket{y} \mapsto \ket{1}\ket{x}\ket{x \oplus y}
\]
Similarly let $\Lambda (U): {\cal B}^{n+1}\rightarrow {\cal B}^{n+1}$
on registers $\cal C$ and $\cal X$ be the controlled-$U$ operation:
\[ \Lambda (U): \ket{0}\ket{x} \mapsto \ket{0}\ket{x} \hspace{1cm}
 \Lambda (U): \ket{1}\ket{x} \mapsto \ket{1}\, U\ket{x}
\]
Let $N$ be the operation of negation in the register $\cal C$.

Suppose that $U\ket{0}=\ket{0}$. Then $\Lambda (U)$ can be
implemented as follows. In addition to the $n-$qubit register $\cal X$
of $U$ we introduce a 1--qubit control register $\cal C$ and an
extra $n-$qubit register $\cal Y$. Consider the sequence of operations
(reading from left to right) in which the square brackets denote the
registers to which the operations are applied:
\[ N[{\cal C}]\,\,\, \Lambda (\tau ) [{\cal C},{\cal X}, {\cal Y}]\,\,\,
\Lambda (\tau ) [{\cal C},{\cal Y}, {\cal X}] \,\,\, U[{\cal X}]\,\,\,
\Lambda (\tau ) [{\cal C},{\cal Y}, {\cal X}]\,\,\,
\Lambda (\tau ) [{\cal C},{\cal X}, {\cal Y}]\,\,\, N[{\cal C}]  \]
If $\cal Y$ is initially set to $\ket{0}$ then after these operations
$\cal Y$ will again be $\ket{0}$ and $\Lambda (U)$ will have been effected
on the registers $[{\cal C}, {\cal X}]$.
This is readily seen by a straightforward calculation.
The $\Lambda (\tau ) $ operations on either side of $U[{\cal X}]$
simply serve to swap the states of the registers $\cal X$ and $\cal Y$.
Thus if ${\cal C}$ is $\ket{0}$ the states in $\cal X$ and $\cal Y$ are 
swapped and $U$ is merely applied to $\ket{0}$. If $\cal C$ is $\ket{1}$
then the states are not swapped and $U$ is applied to the original contents
of $\cal X$.

To measure $\phi$ consider the following procedure {\bf PROC}:\\
Start with registers $[{\cal C},{\cal X}]$ in state $\ket{0}\ket{\lambda}$.
Apply $H$ to $\cal C$, then $\Lambda (U)$ to $[
{\cal C},{\cal X}]$, then $H$ to $\cal C$ again. This results in
the following state in $[{\cal C},{\cal X}]$:
\[ \ket{\psi_\lambda}\ket{\lambda} =
\left( \frac{1}{2} (1+\exp{2 \pi i\phi})\ket{0} +
\frac{1}{2} (1-\exp{2 \pi i\phi})\ket{1}\right) \ket{\lambda} \]
Note that the eigenstate in $\cal X$ has not been corrupted and may
be used again.
Finally measure the control register. This will yield 0 or 1
with probability distribution $\cal P$ given by:
\[ p_0 = \frac{1}{2}(1+\cos 2\pi \phi ) \hspace{1cm}
p_1 = \frac{1}{2}(1-\cos 2\pi \phi )
\]

To get the information of $\phi$ we just repeat {\bf PROC} for many
independent control qubits, sampling the distribution $\cal P$
sufficiently many times to get an adequate estimate of $p_0$.
Suppose we apply {\bf PROC} \,$t$ times successively,
starting with $t$ control qubits  and ending in the state
$\ket{\psi_\lambda}\ket{\psi_\lambda}\ldots \ket{\psi_\lambda}\ket{\lambda}$
and then sample $\cal P$\,\, $t$ times. 
Let $y$ be the number of times that outcome ``0'' occurs. Then by
the weak law of large numbers, for any $\delta > 0$
\begin{equation}\label{est}
 \mbox{Prob}\left( |\frac{y}{t}-p_0 |> \delta \right)
\leq \frac{2}{\sqrt{2\pi}}\exp{\left( -\frac{\delta^2 t}{2p_0 p_1}\right)}
\equiv \epsilon
\end{equation}
Thus with $t$ repetitions we can measure $p_0$ (i.e. $\phi$) to precision
$\delta$ with error probability $\epsilon$.
Note that for fixed $\delta$ the error probability $\epsilon$ decreases
exponentially with $t$ i.e. $t=O(\log (1/\epsilon)$ but the precision
$\delta$ (for fixed $\epsilon$) cannot be efficiently improved --
for each extra bit of precision, $\delta \rightarrow \delta /2$ ,
we require $t\rightarrow 4t$ in (\ref{est}) to maintain a constant level of
$\epsilon$. Hence by this direct method,
the number of bits of precision can be improved only by a correspondingly
{\em exponential} increase in computing effort -- $O(4^l)$ steps
for $l$ bits of precision. This is unacceptable.
To get around this difficulty let us suppose that not only $U$ is
efficiently computable (i.e. in poly($n$) steps) but also that:
\begin{equation}\label{ass}
 \mbox{ {\bf Assumption}:  $U^{(2^j)}$ can be computed in poly($j,n$) steps  }
\end{equation}
This assumption is valid in our application of $U$ being
``multiplication by $y$''. $U^{2^j}$ is then ``multiplication by
$y^{2^j}$'' which can be implemented by a sequence of $j$ repeated
squarings, starting with $y$. It will not, however, be valid for a general 
unitary transformation $U$.

Now assuming (\ref{ass}) we can efficiently improve the precision
$\delta$ as follows i.e. obtain $l$ bits of $p_0$ with computing effort
poly($l$).
Note that $\ket{\lambda}$ is an eigenstate of $U^{2^j}$ with
eigenvalue $\exp{( 2\pi i [2^j \phi \mbox{ mod } 1])}$. To obtain $l$ bits
of $\phi$ with error probability $\leq \epsilon$ we measure (as above)
the values of $2^j \phi \mbox{ mod } 1$
for $j=0,\ldots , l-1$,
    to a fixed precision $\delta =
1/8$ with error probability $\leq \epsilon /l$.
Now if we write $\phi$ in binary then $2^j \phi$ has the point shifted $j$
places to the right and ``mod 1'' removes the integer part. Thus
knowing $2^j \phi \mbox{ mod }1$ to $\pm \frac{1}{8}$ gives the first few bits
of  $2^j \phi \mbox{ mod }1$ i.e. bits $j$ and $j+1$ of $\phi$ itself.
Hence we get about $l$ bits of precision of $\phi$. The probability that all
these bits are correct exceeds $(1-\epsilon /l)^l \geq 1-\epsilon$.
This completes the efficient approximation of $\phi$ under the
assumption (\ref{ass}) above.

Generally (as in Kitaev's factoring algorithm) we will not have available
a pure eigenstate of $U$ but instead some superposition
$\sum a_\lambda \ket{\lambda}$. If we apply {\bf PROC} to this state
with $t$ control bits we will obtain $\sum a_\lambda
\ket{\psi_\lambda}\ldots \ket{\psi_\lambda}\ket{\lambda}$ so that
a measurement of the control bits will yield one of the eigenvalues
$\lambda$ with probabilities $|a_\lambda |^2$. i.e. if we trace out
the eigenstate register $\ket{\lambda}$ the $t$ control qubits
are in a mixture of the repeated states
$\ket{\psi_\lambda}\ldots \ket{\psi_\lambda}   $ with probabilities
$|a_\lambda |^2$. Note that we must apply {\bf PROC} $t$ times {\em before}
any measurement of the control qubits is made. Otherwise each successive
measurement will provide information about a different eigenvalue
and finally we will only obtain information about the {\em average}
value of the $\lambda$'s weighted by $|a_\lambda |^2$,
rather than about some {\em one} of the $\lambda$'s.

In most cases the eigenvalues $exp{(2\pi i \phi)}$ will have {\em rational}
values of $\phi$, $\phi =a/b$. This is because the $U$'s of interest
will have finite order i.e. $U^m = I$ for some $m$ so that $\phi = k/m$ for
some $k$. For example if $U$ is ``multiplication by $y$'' then $U^r =I$
so that $\phi$ must have the form $k/r$ (as noted previously).
In this situation we can find $\phi$ exactly, rather than
just approximately, by choosing a suitably high precision $\delta$.
The minimum separation between any two rational numbers
with denominators $r$ is $1/r$ so we can get $\phi = k/r$ exactly by
measuring it to precision $1/2r >1/2N$ i.e. $1+\log N$ bits. 

Thus we obtain an efficient factoring algorithm based on the novel idea
of determining an eigenvalue of a given simple unitary operation.
Some of the formalism of Kitaev's algorithm may be related to
Shor's method by using the decomposition of the Fourier transform given
in \cite{gr} but it would be interesting to consider 
other problems that might be 
formulatable in terms of the determination of eigenvalues.\\[3mm]
{\large\bf Conclusions}

We have seen that the principal known quantum algorithms all revolve 
around one essential construction, that of the Fourier transform
on an Abelian group. Furthermore the quantum computational speedup
provided by these algorithms may be attributed to 
(non-classical) properties of entanglement operating within the
implementation of classical fast Fourier transform algorithms 
on a quantum computer \cite{J1}. Clearly it would be of great interest to have
other basic ingredients for the construction of new quantum algorithms.
Kitaev's formalism \cite{KIT} as we have illustrated, appears to involve 
such an ingredient. The mathematical construction of the Fourier transform
also extends to non-Abelian groups and it would be interesting to investigate 
problems which can be formulated in terms of  non-Abelian 
Fourier transforms  and the possiblity of their implementation on a
quantum computer. This line of development has also been advocated by 
Hoyer \cite{HO}.
\\[3mm]
{\large\bf Acknowledgements}

The author is grateful for the hospitality of the Institute for
Theoretical Physics, University of California, Santa Barbara
during the Quantum Computers and Quantum Coherence program
during which time a preliminary version of this work was completed.
This work was supported in part by the National Science Foundation
under Grant No. PHY94-07194 and by the Royal Society, London.

\end{document}